\documentstyle[epsf,times]{mn}
\newcommand{\etal}{{et al}\/.}
\begin{document}
\title[{\it Chandra} detection of 3C\,123]{A {\it Chandra} detection of the radio hotspot of 3C\,123}
\author[M.J.~Hardcastle \etal]{M.J.\ Hardcastle, M.\ Birkinshaw and
D.M.\ Worrall\\
Department of Physics, University of Bristol, Tyndall Avenue,
Bristol BS8 1TL}
\maketitle
\begin{abstract}
{\it Chandra X-ray Observatory} observations of the powerful, peculiar
radio galaxy 3C\,123 have resulted in an X-ray detection of the bright
eastern hotspot, with a 1-keV flux density of $\sim 5$ nJy. The X-ray
flux and spectrum of the hotspot are consistent with the X-rays being
inverse-Compton scattering of radio synchrotron photons by the
population of electrons responsible for the radio emission
(`synchrotron self-Compton emission') if the magnetic fields in the
hotspot are close to their equipartition values. 3C\,123 is thus the
third radio galaxy to show X-ray emission from a hotspot which is
consistent with being in equipartition. {\it Chandra} also detects emission
from a moderately rich cluster surrounding 3C\,123, with $L_X (2--10{\
\rm keV}) = 2 \times 10^{44}$ ergs s$^{-1}$ and $kT \sim 5$ keV, and
absorbed emission from the active nucleus, with an inferred intrinsic
column density of $1.7 \times 10^{22}$ cm$^{-2}$ and an intrinsic
2--10 keV luminosity of $10^{44}$ ergs s$^{-1}$.
\end{abstract}
\begin{keywords}
galaxies: active -- X-rays: galaxies -- galaxies: individual: 3C\,123
-- radiation mechanisms: non-thermal
\end{keywords}

\section{Introduction}

The magnetic field strengths in the extended components of
extragalactic radio sources cannot be inferred directly from
observations of synchrotron emission, and so the energy densities and
pressures in the radio-emitting components are poorly constrained. In
order to make progress it is common to estimate `minimum energy' field
strengths (Burbidge 1956), which minimise the energy density required
for a given synchrotron emissivity. This is roughly equivalent
to the assumption that that magnetic and relativistic particle
energy densities are equal (`equipartition'). But without measurements
of magnetic field strengths these assumptions, for which there is no
physical justification, may underestimate the true energy densities by
arbitrary factors.

The magnetic field strength may be measured by observations of the
`synchrotron self-Compton' (SSC) process, in which the
synchrotron-emitting electrons inverse-Compton scatter synchrotron
photons up to X-ray energies. Because the emissivity from this process
depends on the photon number density (which is known from radio
observations) and the electron number density as a function of energy,
observations of SSC emission allow the electron number density to be
inferred, and so determine the magnetic field strength. Such tests
require observations of regions with well-measured volume and a
well-defined synchrotron spectrum with a high photon energy density;
these conditions exist in the hotspots of FRII radio sources. Direct
evidence supporting the equipartition/minimum energy assumptions in
hotspots has come from only two X-ray observations. Harris, Carilli \&
Perley (1994) detected the hotspots of the powerful FRII Cygnus A
(3C\,405) with {\it ROSAT} and showed that the X-ray emission could be
interpreted as being due to the SSC process, with a magnetic field
strength consistent with the equipartition model. [This result was
recently confirmed with {\it Chandra} by Wilson, Young \& Shopbell
(2000).] {\it ROSAT} was not sensitive enough to detect any other SSC
hotspots, though it was used to put lower limits on the field
strengths in some sources (Hardcastle, Birkinshaw \& Worrall
1998). More recently, {\it Chandra} verification observations have
detected the hotspots of 3C\,295, another powerful radio galaxy, at a
level which implies field strengths fairly close to the equipartition
values if the emission process is SSC (Harris \etal\ 2000). Here we
report a third detection, of the E hotspot of the radio galaxy
3C\,123, based on our {\it Chandra} AO1 guest observer (GO)
observations.

3C\,123 is a $z = 0.2177$ radio galaxy, notable for its peculiar radio
structure. Like normal classical double sources it has twin hot spots
on either side of the active nucleus, but the lobes take the form of
diffuse twisted plumes unlike those in any other well-studied object
(e.g. Riley \& Pooley 1978; Hardcastle \etal\ 1997, hereafter
H97). Like Cygnus A and 3C\,295, its radio luminosity is unusually
high for its redshift. For our purposes, its most important feature is
its bright eastern double hotspot. With a flux density of $\sim 6$ Jy
at 5 GHz, it is the second brightest hotspot complex known (after
Cygnus A). The hotspots' structure and synchrotron spectrum are well
known (H97; Meisenheimer \etal\ 1989; Meisenheimer, Yates \& R\"oser
1997; Looney \& Hardcastle, 2000).

Throughout this letter we use $H_0 = 50$ km s$^{-1}$ Mpc$^{-1}$ and
$q_0 = 0$. At the redshift of 3C\,123, 1 arcsec corresponds to
4.74 kpc.

\section{Observations}

\begin{figure}
\epsfxsize 8.4cm
\epsfbox{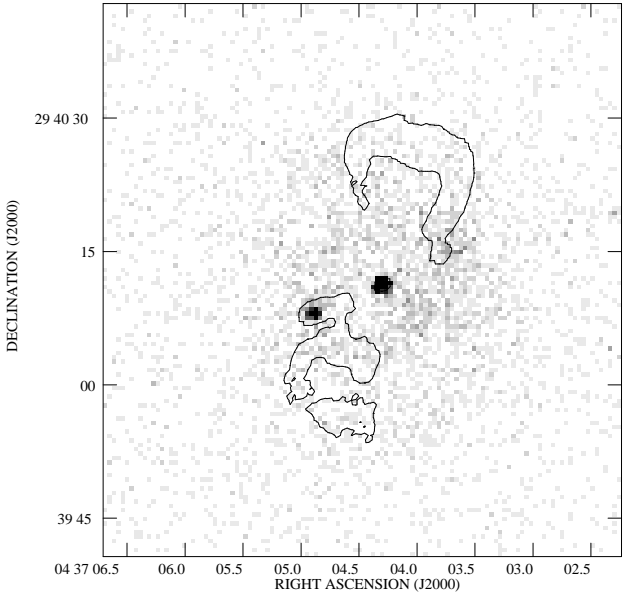}
\caption{Exposure-corrected 0.5--7.0 keV {\it Chandra} image of the
central region of 3C\,123. Linear greyscale: black is $4 \times
10^{-7}$ photons cm$^{-2}$ s$^{-1}$ per standard Chandra pixel (0.492
arcsec on a side). Superposed is the 3 mJy beam$^{-1}$ contour from an
8.4-GHz Very Large Array (VLA) map with 0.6-arcsec resolution (H97),
showing the position of the radio lobes. As discussed in section
\ref{nucleus}, the radio map has been shifted by $\sim 3$ arcsec so
that the radio core aligns with the X-ray nucleus.}
\label{image}
\end{figure}

We observed 3C\,123 with the {\it Chandra X-ray Observatory} for 46.7
ks on 2000 March 21. The source was near the aim point for the S3 ACIS
chip. After filtering for intervals of high background, the usable
exposure time was 38.5 ks. We considered events in the energy range
0.5--7.0 keV, as the spectral response of the instrument is uncertain outside
this range. Fig.\ \ref{image} shows the exposure-corrected {\it
Chandra} image of 3C\,123 in this band. Diffuse cluster emission, an
X-ray nucleus and the eastern hotspot are all detected in X-rays. We
discuss each component in turn. In each case, spectra were extracted
using {\sc ciao}, with the best available responses being constructed for
each extraction region, and analysed using {\sc xspec}. Spectra were
binned such that every bin had $>20$ net counts.

\section{The cluster}

The X-ray counts from 3C\,123 are dominated by diffuse cluster-scale
emission, which is detectable above the background more than an
arcminute away from the central source. The source was known to be
extended from {\it ROSAT} images (Hardcastle \& Worrall 1999), and
there are many faint optical objects in the field of 3C\,123 which may
be cluster members (e.g.\ Longair \& Gunn 1975), though strong
galactic reddening (the source is at $b \approx -12^\circ$, and see
below) means that this cluster has not been studied in detail in the
optical. We detect around 5,000 counts in a 75-arcsec radius circle centred
on the nucleus, excluding the core and hotspot. The overall spectrum
of this region is well fitted with an absorbed MEKAL spectrum with $kT
= 5.0^{+0.6}_{-0.4}$ keV, abundance of $0.47^{+0.12}_{-0.11}$ solar
and a galactic hydrogen column density of $4.3_{-0.3}^{+0.2} \times
10^{21}$ cm$^{-2}$ (errors are $1\sigma$ for one interesting
parameter). 3C\,123 lies behind a well-known molecular cloud system in
Taurus (e.g.\ Ungerechts \& Thaddeus 1987), and our derived column
density is consistent with the total hydrogen column inferred
from radio observations of the molecular cloud system. From HI
absorption against 3C\,123, Colgan, Salpeter \& Terzian (1988) derive
$N_{\rm HI} = 1.97 \times 10^{21}$ cm$^{-2}$ at the velocity of the
cloud system, while the molecular hydrogen column can be inferred from
CO measurements [$W_{\rm CO} \approx 10$ K km s$^{-1}$, Megeath
(private communication)] to be $N_{\rm H_2} \approx 1.6 \times
10^{21}$ cm$^{-2}$, with a large systematic uncertainty [we use a
recent estimate of the conversion factor averaged over the galactic
plane, due to Hunter \etal\ (1997), but this may not be appropriate
for the Taurus region]. We adopt a galactic absorption column of $4.3
\times 10^{21}$ cm$^{-2}$, unless otherwise stated, from now on.

The X-ray spectral fit implies a rest-frame 2--10 keV luminosity from
the cluster within a radius of 75 arcsec (360 kpc) of $2 \times
10^{44}$ ergs s$^{-1}$, consistent with the fitted temperature on the
temperature-luminosity relation (determined largely for Abell
clusters) of David \etal\ (1993). Fig.\ \ref{image} shows a plateau
of X-ray emission on scales comparable to those of the radio source,
with clear structure in the emission (note particularly X-ray
voids to the E and SW of the nucleus) although there is no evidence
for interaction between the X-ray gas and the radio lobes. The voids
may represent large-scale inhomogeneity in the cluster gas; if so,
they would help to explain the peculiar radio structure.  The central
cooling time is a few $\times 10^{9}$ years, so we might expect to see
a cooling flow around the source. But there is no strong evidence for
cooling in the temperature fits; the best-fitting temperature
for the material within 15 arcsec of the nucleus (fixing the abundance
to the value obtained for the whole cluster) is $4.4\pm 0.3$
keV. Since gas with temperatures below $\sim 3$ keV is not observed
even in the centres of well-studied cooling flows (e.g.\ Fabian \etal\
2000, Peterson \etal\ 2000) this is perhaps not surprising.
Our preliminary analysis implies particle densities around the lobes which
are similar to those reported by Hardcastle \& Worrall (2000) using
{\it ROSAT} data, and the measured temperature implies comparable, but
slightly larger, external pressures.

\section{The nucleus}
\label{nucleus}

The point-like nucleus contains $517 \pm 36$ 0.5--7.0 keV counts,
measured in a 2.5-arcsec region about the centroid, with the
background being taken from a 3--5 arcsec concentric annulus. Pileup
is not significant. The X-ray core position (J2000.0) is measured to
be 04 37 04.30 +29 40 11.2. The core position measured from the VLA
radio map of H97 is 04 37 04.375 +29 40 13.86, and this is expected to
be accurate to within about 0.05 arcsec. The X-ray core position is
therefore offset from the (true) radio position by about 3 arcsec. We
attribute this to uncertainties in aspect determination in the
early version of the pipeline software (R4CU4UPD7.4) used to process
the {\it Chandra} data (see URL:
$<$http://asc.harvard.edu/mta/ASPECT/$>$). In Fig.\ \ref{image} we
have aligned the radio data with the X-ray core.

\begin{figure}
\epsfxsize 8.4cm
\epsfbox{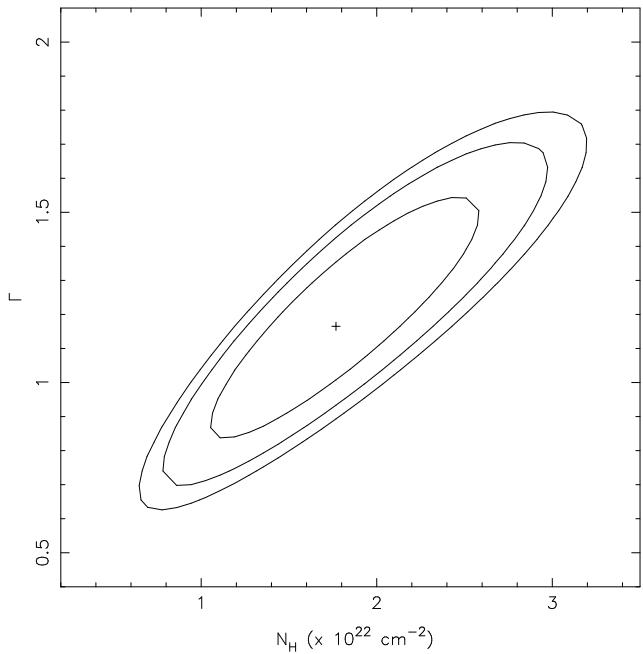}
\caption{Joint confidence contours for the spectrum of the nucleus of
3C\,123 using the model described in the text. Intrinsic absorbing
column is plotted on the $x$-axis, power-law photon index on the
$y$-axis. The contours are at 1, 2 and 3$\sigma$ for two interesting
parameters. The cross marks the best-fit values.}
\label{speccont}
\end{figure}

The spectrum of the nucleus is well fitted with an absorbed,
flat-spectrum power law model. Fitting with free galactic absorption,
the best-fit values of photon index $\Gamma$ and $N_{\rm H}$ are $1.16
\pm 0.14$ and $1.48_{-0.24}^{+0.32} \times 10^{22}$ cm$^{-2}$,
respectively. If we fix galactic absorption at the value derived from
the cluster fits and require the absorber to be at the redshift of the
galaxy, the best-fit $N_{\rm H}$ for the intrinsic absorber is
$1.69_{-0.41}^{+0.51} \times 10^{22}$ cm$^{-2}$, with the photon index
unchanged. This implies a rest-frame 2--10 keV luminosity (assumed
isotropic) of $10^{44}$ ergs s$^{-1}$, comparable to the luminosity of
the nuclear component in 3C\,295 (Harris \etal\ 2000). Fits in which
the absorbing column is constrained to the galactic value are much
poorer and require an inverted nuclear spectrum. The column density
required for the intrinsic absorber is considerably lower than the $4 \times
10^{23}$ cm$^{-2}$ inferred for Cygnus A by Ueno \etal\ (1994), and
more comparable to that inferred for the much lower luminosity nucleus
of Hydra A by Sambruna \etal\ (2000). The best-fit photon index is
flatter than might be expected from, for example, the photon indices
of radio-loud quasars (e.g.\ Lawson \& Turner 1997) or other radio
galaxies (Sambruna, Eracleous \& Mushotzky 1999) although the errors
are large; as shown in Fig.\ \ref{speccont}, more reasonable values of
$\Gamma$ are allowed in conjunction with somewhat higher intrinsic absorbing
columns.  The inferred absorbing column in front of the nucleus
explains the non-detection of this core component in the {\it ROSAT}
HRI image (Hardcastle \& Worrall 1999).

\section{The hotspot}

The E hotspot complex is detected with $145 \pm 32$ 0.5--7.0 keV
counts, using a 2.5-arcsec source circle and concentric 3--5 arcsec
background annulus.  The X-ray hotspot is positionally coincident with
the larger, `secondary' hotspot of the eastern hotspot pair in the
radio (after the X-ray and radio cores have been aligned). The X-ray
emission appears to be slightly elongated in an east-west direction,
matching the radio. (The X-ray structure will be discussed in more
detail in a subsequent paper which will include results of further
radio observations now in progress.)  The X-ray spectrum of the
hotspot is well fitted with a power law with $\Gamma = 1.6 \pm 0.3$,
with the absorbing column fixed at the galactic value. The
corresponding unabsorbed 1-keV flux density is $4.6 \pm 0.9$ nJy.

\begin{figure}
\epsfxsize 8.4cm
\epsfbox{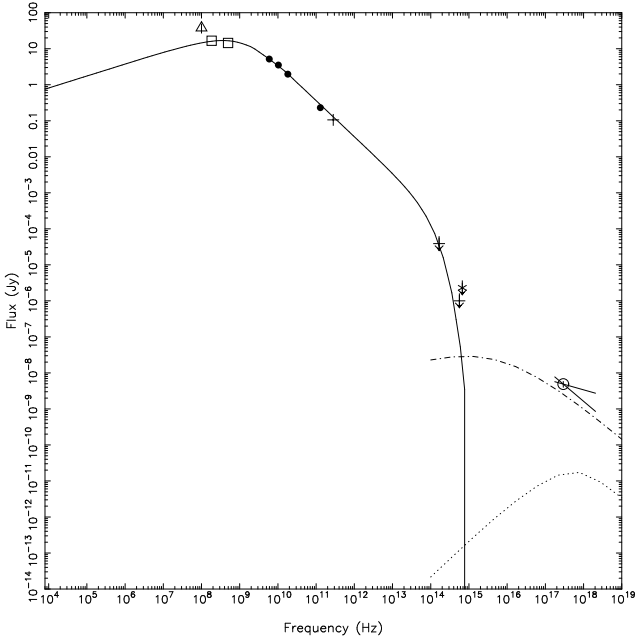}
\caption{The spectrum of the secondary hotspot of 3C\,123 at
equipartition. Points are from data as described in the text. Symbols
indicate the source of the data, as follows: filled circles, Looney
\& Hardcastle (2000); crosses, Meisenheimer \etal\ (1989);
squares, Stephens (1987); triangle, Readhead \& Hewish (1974); star,
optical limit from HST data; open circle, {\it Chandra} data
point. Arrows denote an 
upper limit. Error bars are smaller than the symbols
in most cases. The
solid line is the model synchrotron spectrum. The dash-dotted line shows
the predicted SSC spectrum at equipartition and the dotted line shows
inverse-Compton emission due to scattering of microwave background
photons. The two solid bars through the X-ray data point show the 0.5--7.0
keV band of the {\it Chandra} data and the $1\sigma$ range of photon
indices permitted by the data. Frequencies are plotted in the
rest frame of the radio source.}
\label{flux}
\end{figure}

We used the synchrotron-self-Compton code described by Hardcastle
\etal\ (1998) to predict the SSC flux density expected at this
frequency from the hotspots. The basic model for the hotspots is
described by Looney \& Hardcastle (2000). The larger hotspot is
treated as a cylinder of $1.14 \times 0.54$ arcsec (length times
radius); the fainter `primary' (more compact) hotspot is a cylinder of
$0.74 \times 0.14$ arcsec, based on the MERLIN maps of H97. Radio flux
densities of the two components are taken from Looney \&
Hardcastle. In addition to these, we have used infra-red and optical
upper limits and a 231-GHz data point from Meisenheimer \etal\ (1989,
1997) and archival HST observations, and low-frequency radio data from
Readhead \& Hewish (1974) and Stephens (1987). As these data do not
resolve the two hotspot components, we have {\it approximately}
corrected them by scaling by the appropriate factors measured from the
5-GHz data. Looney \& Hardcastle showed that the radio-to-mm spectra
of the two hotspots are well modelled as broken power laws, and we
adopt the break energies they found. The apparent low-frequency
turnover in the spectrum observed by Stephens (1987) requires a
low-energy cutoff in the electron energy spectrum corresponding at
equipartition to a minimum Lorentz factor $\gamma_{\rm min} \approx
1000$, and we adopt this value, although Stephens' flux densities are
inconsistent with a larger flux at a lower frequency derived from the
scintillation measurements of Readhead \& Hewish. If we were to adopt
the scintillation measurements as our low-frequency constraint, we
would obtain $\gamma_{\rm min} \approx 400$, which is more consistent
with the value inferred for the hotspots of Cygnus A by Carilli \etal\
(1991); but this would not significantly affect our
conclusions. (Scheduled low-frequency VLBA observations should give a
definitive answer.) An upper limit on the maximum Lorentz factor is
given by the non-detection in the IR, $\gamma_{\rm max} < 3.6 \times
10^5$; a lower limit is given by the detection at 231 GHz,
$\gamma_{\rm max} > 8 \times 10^4$. The SSC emissivity at 1 keV turns
out to be insensitive to $\gamma_{\rm max}$ if it lies between these
two values, and so we fix $\gamma_{\rm max}$ at its largest
value. With these parameters, the equipartition field strengths of the
two eastern hotspot components, assuming no contribution to the energy
density from non-radiating particles such as protons, are 24 nT
(primary) and 16 nT (secondary), and the predicted SSC flux densities
at 1 keV are respectively 0.44 and 2.6 nJy. The predicted photon index
at this frequency is 1.55 (of course, this is simply a function of the
electron energy spectrum, and so is true for any inverse-Compton
process). Fig.\ \ref{flux} shows the synchrotron fluxes and SSC
prediction for the secondary hotspot. The predicted SSC flux density
for the much weaker western hotspot pair is negligible, $\sim 0.07$
nJy, corresponding to 2 {\it Chandra} counts in this observation.

These predictions are relatively insensitive to cosmological
parameters; for example, using a cosmology where $\Omega_{\rm matter}
= 1.0$ gives a 3 per cent decrease in the expected flux density from
the secondary, while using $H_0 = 70$ km s$^{-1}$ Mpc$^{-1}$,
$\Omega_{\rm matter} = 0.3$, $\Omega_\Lambda = 0.7$ gives a 2 per cent
decrease. The hotspots may be projected, but this also has only a weak
effect: if the projection angle is $45^\circ$ then the true long axis
is larger by a factor $\sqrt{2}$ and the predicted flux is about 7 per
cent lower. One clear systematic error in the calculations comes from
an assumption of spherical symmetry for the scattering geometry which
is used in the SSC code. Because a cylinder has a higher surface area
to volume ratio, the mean photon density is lower in a cylinder than
in a sphere for a given volume synchrotron emissivity. We estimate
that this effect is less than 10 per cent for the secondary component,
but we may be overestimating the SSC flux from the fainter primary by
30 per cent (these values are increased if there is substantial
projection).

Comparing our predicted flux densities with the data, we see that the
magnitude of the observed flux density, its origin in the larger
eastern hotspot and its photon index are all in good agreement with
the equipartition predictions of the SSC model, as is the non-detection of the
western hotspot pair. The observed 1-keV flux density of the E
hotspots is somewhat higher than the predicted value for an
equipartition field, though only by about 1.5 standard
deviations. Most of the changes to the model discussed above have the
effect of reducing the SSC flux density; to increase it we must reduce
the magnetic field strength or find an additional (external) source of
photons. If the magnetic field in the secondary hotspot is reduced by
25 per cent to 12 nT, the secondary can produce all the flux seen in
X-rays. Neglecting the small SSC contribution from the primary
hotspot, the {\it Chandra} data with their associated uncertainties
imply within the SSC model that the magnetic field strength in the secondary
hotspot is $12 \pm 2$ nT ($1\sigma$ statistical errors only).

One possible external source of photons is the active nucleus of
3C\,123. Inverse-Compton (IC) scattering of photons from the active
nucleus will make a significant contribution to the X-ray emission if
such photons (at frequencies around $10^{11}$ Hz, since $\gamma_{\rm
min} \approx 1000$) are comparable in number density to the
synchrotron photons. At this frequency, the number density of
synchrotron photons is approximately 0.08 m$^{-3}$ Hz$^{-1}$; since
the hotspots are a projected 37 kpc from the nucleus, a similar number
density would be produced from the nucleus if its luminosity at this
frequency as seen by the hotspot were $\ga 2 \times 10^{27}$ W
Hz$^{-1}$ sr$^{-1}$, which would correspond to a flux density of $\ga
100$ Jy at $10^{11}$ Hz. (This condition is equivalent to $S_{\rm
core} = (3D/2R)^2 S_{\rm hs}$, where $D$ is the core-hotspot distance
and $R$ is the hotspot radius, and the two fluxes are measured at the
required frequency.) The observed core flux density of 3C\,123 at
$10^{11}$ Hz is about 40 mJy, though it may be variable (Looney \&
Hardcastle 2000), so that isotropic radiation from the core cannot
provide the required photon density. However, if the core emission at
this frequency is beamed, the hotspot will see the core as having a
higher luminosity than the one we observe. We assume the most
favourable case for this model of no misalignment between the pc- and
kpc-scale jet, though such good alignment is not often observed, and
we neglect effects due to the finite angle subtended by the hotspot at
the nucleus, which may be significant. The ratio of required to
observed flux, ${\cal R}$ (${\cal R} \approx 2500$ for rough equality
of predicted SSC and IC flux densities), then constrains $\beta$, the
bulk speed in the nucleus, and $\theta$, the angle of the core-hotspot
vector to the line of sight:
\[
{{(1-\beta)^{-(2+\alpha)} + (1+\beta)^{-(2+\alpha)}}\over{(1 -
\beta \cos \theta)^{-(2+\alpha)} + (1 +
\beta \cos \theta)^{-(2+\alpha)}}} =
{(1-\cos \theta)^{-(1+\alpha)}{\cal R}\over{\sin^2\theta}}
\]
where the term $(1-\cos \theta)^{-(1+\alpha)}$ approximately corrects
for the anisotropic nature of the resulting IC emission (e.g.\ Jones,
O'Dell \& Stein 1974), and the core is treated as a two-sided jet with
a power-law spectrum; the $\sin^2\theta$ term incorporates the effects
of projection. If we assume a core spectral index $\alpha = 0.5$, then
we can obtain ${\cal R} \approx 2500$ for plausible $\beta$,
corresponding to bulk Lorentz factors $\sim 4$--$10$, if the source is
within 50 degrees of the plane of the sky.  (In reality, this value
for $\alpha$ is probably an overestimate, since the $10^{11}$ Hz
photons seen by the nucleus will be blueshifted from lower frequencies
where core spectra are typically flat, $\alpha \approx 0$. Lower
$\alpha$ requires higher $\beta$.) As no jet has been detected in
3C\,123 and its optical emission lines are weak, the angle to the line
of sight is not constrained, and so we cannot rule out a contribution
to the X-ray emission from nuclear IC scattering. However, a
contribution at approximately the same level as the SSC emission would
not affect the conclusion that the hotspot is close to equipartition;
in fact, it might account for some of the difference between the
equipartition prediction and the observed 1-keV flux density. But if
nuclear IC emission {\it dominates} the observed X-rays, then the
hotspot could have a magnetic field higher than the equipartition
value, and this cannot be ruled out by the present observations. For
example, ${\cal R} = 2.5 \times 10^4$, corresponding to an IC emissivity ten
times the SSC value, can be obtained for a bulk Lorentz factor
$\approx 6$ if the source is close to the plane of the sky and $\alpha
= 0.5$, and would require $B \approx 4B_{\rm eq}$. Such a model requires a
coincidence to explain the similarity of the observed emissivity to
that predicted by the simple SSC model with a near-equipartition
field.

In carrying out the SSC calculations we have assumed that the hotspots
are homogeneous, that they contain no relativistic protons, and that
the small-scale filling factor is unity. From the MERLIN maps of H97
we know that the hotspots do have internal structure on 100-pc scales,
although the variations in surface brightness are not very large;
there is no evidence for filamentary structures of the kind seen in
radio lobes. As discussed in Hardcastle \& Worrall
(2000), the general effect of a filling factor less than unity is to
increase the SSC emissivity, but the results are dependent on the
geometry of the synchrotron-emitting regions, particularly if the low
filling factor is a result of a spatial variation in electron density
in a relatively constant magnetic field. Our present results may be
taken as evidence against low filling factors in the hotspots, as such
filling factors would require coincidences to produce X-ray emission
at the observed levels. Similarly, if the particle population is
energetically dominated by non-synchrotron-emitting particles such as
relativistic protons, it is a coincidence that the energy density in
magnetic field corresponds so closely to that in the
synchrotron-emitting electrons.

Although SSC emission is a required process, we cannot rule out the
possibility that the magnetic field strength is much greater than the
equipartition value and that some other process happens to produce
X-ray emission at a level consistent with the SSC model. One such
process which we have already discussed is the IC
scattering of nuclear photons. Some other simple models can be
rejected. Thermal bremsstrahlung can be ruled out by the compact
structure seen with {\it Chandra} and the flat X-ray spectrum, while,
as shown in Fig.\ \ref{flux}, IC scattering of the
microwave background is two orders of magnitude too weak to be
responsible for the observed emission. But more speculative models
remain possible.  For example, we cannot rule out the possibility that
the X-ray emission is synchrotron radiation from an arbitrarily chosen
population of electrons. This model has been invoked to explain some
other X-ray hotspots (e.g. Harris \etal\ 1999) in which the
IC or SSC models do not seem to work
well. In the case of 3C\,123, it requires a coincidence to explain the
close similarity between the observed X-ray emission and the
predictions of the SSC model.

One model which makes quantitative predictions about the origin and
properties of this second population of electrons is the
proton-induced cascade (PIC) model of Mannheim, Kr\"ulls \& Biermann
(1991). In this model, the high-energy electrons which produce X-ray
synchrotron emission are the end result of photomeson production on a
population of ultra-relativistic protons, through pion decay and pair
production. If protons are present in the jet, they should undergo
shock acceleration in the hotspot; there is some evidence for
high-energy protons in the lobes of FRII sources (Leahy \& Gizani
1999, Hardcastle \& Worrall 2000). Mannheim \etal\ considered the
hotspot of 3C\,123 and predicted that the PIC process would dominate
over SSC if protons were highly energetically dominant in the hotspot
and in equipartition with magnetic fields at the level inferred by
Meisenheimer \etal\ (1989), somewhat higher than our equipartition
fields. Their predicted X-ray flux at 1 keV is $\sim 30$ nJy, nearly a
factor 10 higher than the observed value, so a model with extreme
proton dominance does not seem to be consistent with the
data. However, if protons have energy densities closer to those of the
electrons, we cannot rule out an origin from PIC-generated electrons
for some or all of the observed X-rays; the spectra of the two
processes are not distinguishable with our data. Once again, though, a
moderate contribution from the PIC process would not affect our
conclusions regarding the closeness of the hotspot to equipartition,
while a model in which PIC was responsible for {\it all} the emission
would require fine-tuning of the energy fraction in fields, electrons
and protons, and so seems less plausible than the simple SSC model.

\section{Conclusions}

The X-ray emission from the E hotspot of 3C\,123 is consistent with a
SSC model, with the inferred magnetic field strength close to the
value predicted from equipartition of energy between the magnetic
field and the synchrotron-emitting electrons. Other models are
possible, but require coincidences to explain the closeness of the
X-ray flux density and (in some cases) the photon index to the SSC
prediction. This reinforces a conclusion already drawn from
observations of Cygnus A and 3C\,295 that the magnetic field strengths
in typical hotspots are near their equipartition values. Although
there is still no {\it a priori} reason to expect equipartition, it is
now very likely that it is achieved in the hotspots of at least some
fraction of the source population.  Observations of inverse-Compton
scattering of microwave background photons from radio lobes may tell a
similar story (Feigelson \etal\ 1995; Tsakiris \etal\ 1996; Tashiro
\etal\ 1998). Even the deviations from equipartition reported by
Tashiro \etal\ require a magnetic field strength only a factor $\sim
2$ below the equipartition value. However, some X-ray detections of
hotspots (e.g. 3C\,120, Harris et al 1999; Pictor A, Wilson, Young \&
Shopbell 2001) are at a level much too bright to be consistent either
with synchrotron emission (from the electron population responsible
for the radio and optical synchrotron radiation) or with SSC at
equipartition. More observations are necessary to demonstrate that
Cygnus A, 3C\,295 and 3C\,123 are typical of most radio
sources. We will report on the results of scheduled {\it Chandra}
observations of further hotspot sources in a future paper.

\section*{Acknowledgements}
We thank all those involved with the design and operation of {\it
Chandra} for doing such an excellent job, and particularly the staff
of the {\it Chandra} X-ray Center for their help with data
analysis. We are grateful to an anonymous referee for a careful
reading of the paper and constructive suggestions. The National Radio
Astronomy Observatory Very Large Array is operated by Associated
Universities Inc., under co-operative agreement with the National
Science Foundation.

\bsp

\end{document}